\documentclass[twocolumn,runningheads]{svjour3}% running heads, twocolumns
\usepackage{color,colortbl}
\definecolor{Gray}{gray}{0.90}

\usepackage{graphicx}
\usepackage{epstopdf}
\usepackage{mathptmx}
\usepackage[authoryear]{natbib}
\bibpunct{(}{)}{;}{a}{,}{,}
\journalname{Experiments in Fluids}
\newcommand{\x}[2]{\mbox{#1\hspace{0.1em}\raisebox{0.13ex}{x}\hspace{0.1em}#2}}

\pdfoutput=1

\begin{document}

\title{Velocity measurements in the near field of a diesel fuel injector by
ultrafast imagery}
%\subtitle{}

\titlerunning{Velocity in the near field of a diesel spray}

\author{David Sedarsky	\and
        Sa\"id Idlahcen \and
        Claude Roz\'e   \and
        Jean-Bernard Blaisot}


\institute{D. Sedarsky \and S. Idlahcen \and C. Roz\'e \and J-B. Blaisot \at
              UMR 6614 CORIA, CNRS, Universit\'e et INSA de Rouen\\
              BP 12, avenue de l'Universit\'e\\
              76801 Saint Etienne du Rouvray Cedex, France\\
              Tel.: +33(0)23-295-3691\\
              Fax: +33(0)23-291-0485\\
              \email{david.sedarsky@coria.fr}%  \\
              %\emph{Present address:} of F. Author
}

\date{Received: 22 July 2012/Revised: 27 November 2012/Accepted: 21 December 2012}
\maketitle

\begin{abstract}
This paper examines the velocity profile of fuel issuing from a high pressure
single-orifice diesel injector. Velocities of liquid structures were determined
from time-resolved ultrafast shadow images, formed by an amplified two-pulse
laser source coupled to a double-frame camera. A statistical analysis of the
data over many injection events was undertaken to map velocities related to
spray formation near the nozzle outlet as a function of time after start of
injection (SOI).
These results reveal a strong asymmetry in the liquid profile of the test
injector, with distinct fast and slow regions on opposite sides of the orifice.
Differences of $\sim$100~m/s can be observed between the `fast' and `slow' sides
of the jet, resulting in different atomization conditions across the spray. On
average, droplets are dispersed at a greater distance from the nozzle on the
`fast' side of the flow, and distinct macrostructure can be observed under the
asymmetric velocity conditions. 
The changes in structural velocity and atomization behavior resemble flow
structures which are often observed in the presence of string cavitation
produced under controlled conditions in scaled, transparent test nozzles.
These observations suggest that widely-used common-rail supply configurations
and modern injectors can potentially generate asymmetric interior flows which
strongly influence diesel spray morphology.
The velocimetry measurements presented in this work represent an effective and
relatively straightforward approach to identify deviant flow behavior in real
diesel sprays, providing new spatially resolved information on fluid structure
and flow characteristics within the shear-layers on the jet periphery.

\end{abstract}

\section{Introduction}
\label{intro}
At present, there are a number of well-known physical phenomena in spray flows
which are not fully understood, in the sense that their complete behavior cannot
be predicted. High pressure injection used to atomize liquid fuel in combustion
applications is a prime example. The full atomization process results in a
droplet granularity which cannot be fully controlled by injection parameters,
e.g.~injection pressure, nozzle geometry, etc. \citep{Shavit1995}.
In the case of diesel fuel injection, interactions between different parameters
on a variety of scales complicate a full description of the injection process.
These difficulties include: small orifice diameters ($\sim$100~$\mu$m), high
injection pressures ($\sim$200~MPa), large optical depths, and short injection
durations.
Recent work in direct numerical simulation (DNS) applied to liquid injection
phenomena has demonstrated that realistic simulations of spray events are
possible \citep{Menard2007,Lebas2009}, albeit at the cost of large computation
times. However, the validation of numerical models with experimental results
remains a challenge. Good quality visualizations of diesel fuel sprays may be
found in the literature, though spray regions with large optical depth
(\textit{OD}) are typically under-resolved or inaccessible. Recently, a number
of innovative visualization techniques have been developed to address the
difficulties presented by multiple scattering in sprays, e.g. ballistic imaging
\citep{Sedarsky2006,Sedarsky2009,Sedarsky2011,Linne2009,Idlahcen2012}, X-ray
diagnostics \citep{Ramirez2009}, and background reduction schemes
\citep{Kristensson2010}.

While time-resolved images can capture important information related to the
atomization process, the characterization of the spray must include the
spatially-resolved instantaneous velocity to fully describe the injection
features. Measurements of near-nozzle spray regions which include such
information can be used to track the kinematics of spray formation, revealing the
inception and growth of small instabilities which grow to dominate the
downstream spray behavior and drive the atomization process. In turn, this
information can serve to validate numerical simulations of primary breakup and
spray morphology.

The near-nozzle regions of practical sprays are often easily disturbed,
precluding the use of diagnostics which perturb the flow conditions.
Measurements of liquid velocities in diesel sprays are further complicated by
the small relative scale of the spray features compared to the magnitude of the
flow velocity (on the order of $\sim$500~m/s). Although a number of optical
techniques are routinely applied to track fluid motion at this scale and
magnitude, most are unsuitable for application in the vicinity of a dense stream
of fuel. This is due in large part to the numerous scattering interactions with
droplets and other liquid structures in the flow which attenuate and redirect
significant portions of the optical signal.

\subsection{Velocity methods}
%\paragraph{Paragraph headings} Use paragraph headings as needed.
Single point techniques, such as laser Doppler velocimetry (LDV) and related
approaches \citep{Bachalo1994}, can provide velocity information for specific
features of the spray but care must be taken to account for errors in
applications where scattering effects cannot be neglected. In general, it is
not practical to apply these diagnostics in regions with significant scattering.
Moreover, LDV measurements are based on the properties of spherical droplets,
and therefore unsuited to the primary atomization regions of high-pressure
sprays, where the shape of the liquid structures is often complex.

Laser correlation velocimetry (LCV) is a promising single-point technique which
may be applicable to sprays with moderate to high \textit{OD}
\citep{Chaves2004}. However, interpretation of the LCV information is not
straightforward. At present, the technique is suited to monitor spray
performance under some conditions, but may be difficult to apply for spray
characterization \citep{Hespel2012}.

Particle image velocimetry (PIV) is the preferred technique for velocity
imaging, due to the accuracy with which it can be applied, provided that the
region of interest can be seeded with tracer particles that follow the flow and
provide well-resolved image markers \citep{Raffel2007}. Here, correlation
methods are used to extract velocity information from image-pairs or double exposure images
dominated by light scattered from the seed particles, allowing the flow to be
mapped and tracked.

The prospect of seeding the flow in the present work is undesirable, due to the
sensitivity of atomization process. In addition, much of the velocity data
needed to inform our understanding of the spray morphology is concentrated in
the shear layers near the liquid/gas interface as well as within the liquid
features themselves. These non-laminar, multiphase conditions, coupled with the
large density differential between the liquid and the gas make effective seeding
of the flow very difficult. However, the correlation methods applied in PIV can
be adapted for calculating velocity from images of unseeded
flows\citep{Tokumaru1995}.

This approach, generally known as image correlation velocimetry (ICV), uses
matching algorithms to calculate velocity either by tagging the flow with
trackable features \citep{Kruger1999}, matching the motion of naturally
occurring features within the flow \citep{Sedarsky2006}, or by predicatively
morphing and validating the scalar field \citep{Marks2010}.
Here, the form of the measured scalar field and the resulting correlation
topology can vary widely compared to the well-behaved intensities generated by
seed particles.
This difference in the variability and structure of the sampled intensity fields
is the fundamental difference between PIV techniques and ICV. The former
achieves highly accurate matching by correlating fields with generic,
well-separated, and moderately uniform intensity peaks. The latter applies
specialized matching approaches to signals which are not structured
appropriately or lack the signal-to-noise levels necessary to apply standard PIV
analysis.

High-quality results can be obtained with ICV methods, however care
must be taken to validate velocity results as the correlation errors associated
with unseeded images can be appreciably higher than PIV \citep{Fielding2001}.
ICV methods are appropriate for the present work, which requires non-intrusive
temporal- and spatially resolved velocity measurements in the near-field regions
of a diesel spray.

The objective of this article is to examine the flow conditions
of a single-hole diesel fuel injector nozzle.
% , which was designed produce a $1/6$-scale spray
% related to fuel delivery from a 6-hole commercial diesel fuel injector (Renault
% K9K Euro4).
To this end, time-resolved ultrafast shadow images of the
fuel spray were acquired at a series of injection pressures and times following
the start of injection (SOI). Instantaneous velocities of jet structures and
droplets were obtained by matching and validating the motion of liquid-gas
boundary features resolved in shadow images of a high-pressure fuel spray.

\subsection{Resolved feature matching}
Although the spray presents the viewer with a complex three-dimensional mass of
inhomogeneous features along its centerline, the conical symmetry of
plain-orifice jet allows the time-resolved spray periphery to be reliably
interpreted as a two-dimensional measurement region \citep{Sedarsky2012_a}. By
positioning the object plane of the imaging system at the center of the spray,
the structure and shear layers of the spray edges can be spatially resolved in
the limited region bounded by the depth-of-focus of the light collection optics.
In addition, while multiply-scattered light is present in the images, this
arrangement emphasizes measurements in the regions of the spray least affected
by this noise contribution.
The shadow images in this work were generated using a transillumination imaging
system, which is discussed in detail in Section \ref{acquisition}. Source light
for the measurements was supplied by an amplified femtosecond laser system
configured for two-pulse operation, with an imaging arrangement similar to the
system of Sedarsky et al. \citep{Sedarsky2009}. Here, the light collection and
transmission to the image plane result in a depth-of-field of the order of
150~$\mu$m, estimated from the optical parameters as discussed in
\citep{Sedarsky2012_a}. However, since the present work is focused on the spray
periphery, the optical Kerr effect shutter was omitted in the current implementation.

The details of the image processing and matching procedures used to
obtain velocity information are discussed in Section \ref{analysis}, followed by
a presentation of statistical velocity profiles compiled from the single-shot
velocity results.
The velocity profiles given in Section~\ref{injector} provide a convenient
perspective for viewing the relative motion of the jet boundary and ligaments or
droplets within the depth-of-field of the imaging system. This approach
highlights structural instabilities appearing as the liquid exits the orifice,
revealing behavior which influences breakup and the overall morphology of the
spray.
In Section~\ref{atomization}, anomalous behavior of the test injector is
identified in the velocity profiles and related to spray morphology visible in
individual spray images. Finally, the probable sources of the anomalous breakup
modes and spray structure are discussed in light of these velocity results. 

%  Section \ref{injector} presents a compilation of
% results from this analysis, statistically mapping velocity throughout the near
% field region of the injector nozzle for different orientations and injection
% pressures. Section \ref{atomization} discusses the spray structure and describes
% the atomization process in light of these results.

\section{Image acquisition}
\label{acquisition}

\begin{figure*}[bht]
\centering
  \begin{minipage}[b]{0.65\linewidth}
    \includegraphics[scale=1.4]{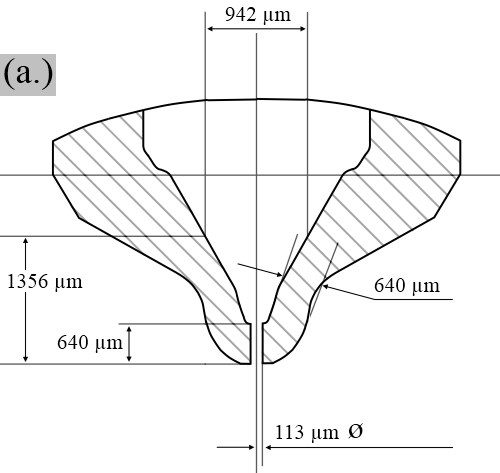}\hfill
    \includegraphics[scale=3.5]{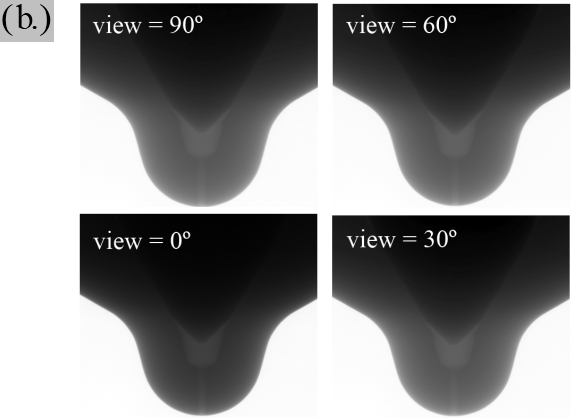}\vfill
  \end{minipage}
  \caption{Detailed views of the diesel test injector nozzle: (a)
  Schematic view; (b) X-ray transmission shadowgrams showing internal geometry at
  0, 30, 60, and 90$^{\circ}$.}
  \label{fig:injector}
\end{figure*}

Sets of time-correlated images were obtained to analyze the kinematics of a diesel
fuel injection spray. The spray was illuminated by a double-pulse femtosecond
laser system consisting of two synchronized regenerative Ti-Sapphire amplifiers
(Coherent Libra) seeded by a common Ti-Sapphire oscillator (Coherent Vitesse, 80
MHz). The short-pulse light from the oscillator is stretched and separated into
two beams, which are amplified as they traverse the gain medium in their
respective amplifiers. After compression, the two beam paths are recombined to
form a single beam containing pairs of pulses with precisely known time
separations.

By adjusting the selection of oscillator pulses entering the
amplifiers, the delay between consecutive output pulses can be adjusted from
$12.5\pm2$ ns to $500 \mu$s. Here, the adjustment step corresponds to the period
of the oscillator, and the variability in the adjustment stems from the possible
optical path differences between selected oscillator pulses.

The resulting source light is a 1~kHz train of 100~fs pulse-pairs, confined to a
low divergence beam with a 10~mm diameter, a center wavelength of 800~nm, and an
average energy per pulse of 3.7~mJ. The oscillator, amplifiers and beam
combination optics are carefully adjusted to obtain similar pulses in each pair,
in terms of pulse duration, amplitude, optical alignment, and polarization.

This two pulse beam was directed toward the injector which is housed in a vented
(atmospheric pressure) enclosure with optical access to facilitate the study of
relevant fuels without contaminating the system optics. The measurements were
carried out using a calibration liquid (Shell NormaFluid, ISO~4113) with
properties similar to diesel fuel and precisely controlled viscosity, density,
and surface tension specifications (see Table \ref{table:normafluid}).

\begin{table}[h]
% NormaFluid properties from Ndiaye 2012
\caption{Properties of ISO 4113 calibration oil.}
\label{table:normafluid}
\begin{minipage}[b]{0.5\linewidth}\centering
\begin{tabular}{lll}
\hline\noalign{\smallskip}
Density & Viscosity & Surface Tension  \\
\noalign{\smallskip}\hline\noalign{\smallskip}
821 kg/m$^3$&	0.0032 kg/(m$\cdot$s)&	0.02547 N/m \\
\noalign{\smallskip}\hline
\end{tabular}
\end{minipage}
\end{table}

The injector nozzle used in the measurements was a hydro-ground, plain-orifice
test nozzle with a conical micro-sac construction and needle valve closure.
% The injector was designed produce a $1/6$-scale spray related to fuel
% delivery from a 6-hole commercial diesel fuel injector (Renault K9K Euro4).
The injector was designed to produce a single-hole spray related to fuel
delivery from a 6-hole commercial diesel fuel injector. A schematic view of the
nozzle is given in Fig.~\ref{fig:injector}, together with a view of the specific
internal geometry given at 4 different angles by x-ray transmission shadowgrams.
The relevant nozzle and spray parameters are listed in Table
\ref{table:injectionparams}.

\renewcommand*\arraystretch{1.4}
\begin{table}[h]
\caption{Test nozzle and injection parameters.}
\label{table:injectionparams}\hspace{0.5cm}
\begin{minipage}[b]{0.5\linewidth}\centering
  \begin{tabular}{|l |l |}
  \hline\rowcolor{Gray}
  Nozzle properties~~~ & Spray conditions~~~~ \\
  \hline
  ~~P$_{inj} = 40$--$100$~MPa & ~~P$_{back} = 0.1$~MPa \\
  ~~C$_d = 0.84$ & ~~U$_B = 300$--$500$~m/s \\
  ~~K$_N$ $ = 1.002$ & ~~Re $ = 9000$--$14000$ \\
  ~~$\ell/$d $=5.66$& ~~We $ = 1140$--$1800$\\\hline
  \end{tabular}
\end{minipage}
\end{table}
\renewcommand*\arraystretch{1}

\begin{figure}[h]
\centering
  \includegraphics[width=0.4\textwidth]{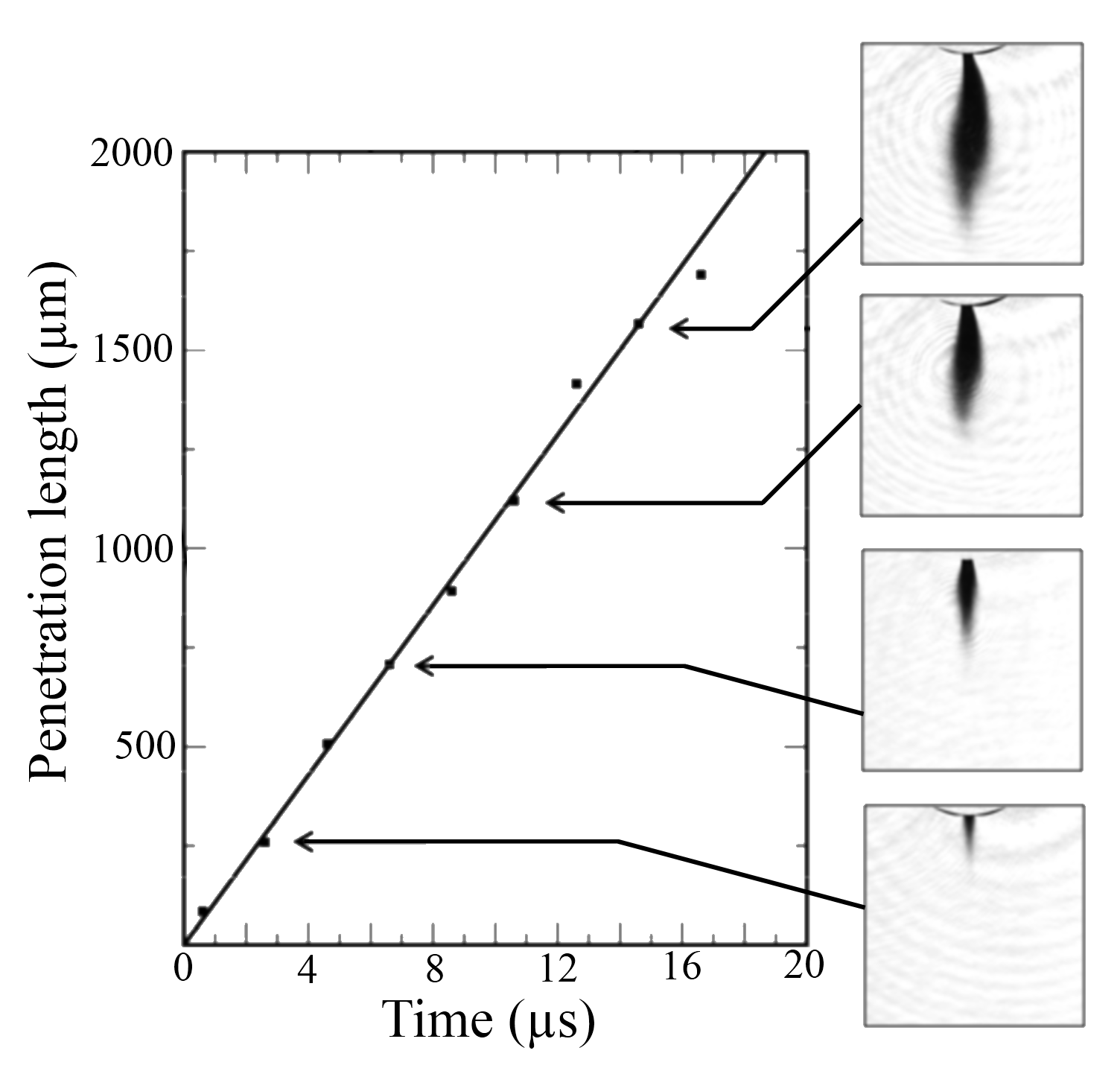}
  \caption{Average jet penetration vs. time. The inset images show the liquid
  jet, averaged over 200 shots. Precise injection timing from SOI in each
  measurement was identified by penetration level, which is a linear function of
  time for early injection times.}
  \label{fig:injectiontiming}
\end{figure}

The injector assembly was mounted on three-way translation stage, and fuel was
delivered to the injector housing by a pump through a Common-Rail accumulator
capable of supplying system pressures up to $\sim$100~MPa.
% The start and duration of each injection event was set by the action of a
% piezo-electric actuator controlling the pressure balance on the needle valve in
% the injector assembly, which allows the system to maintain a constant injection
% pressure throughout the range of needle positions.
The start and duration of each injection event was set by the action of a
balanced servo solenoid mechanism which allows precise control of needle lift.
Each injection event was driven electronically by a staccato-style current pulse
of 400~$\mu$s duration, delivered to the injector at a repetition rate of 1~Hz,
where the reference clock for the complete system was sourced from the laser
oscillator (80~MHz). In this arrangement the onset of each injection event was
subject to timing jitter on the order of 1~$\mu$s, creating a sizeable
systematic error with regard to precise measurement timing relative to SOI.
This uncertainty was reduced by noting that jet penetration length for a high
pressure diesel spray is a linear function of time at early injection times (see
Fig.~\ref{fig:injectiontiming}).
For each separate time delay case, the hydrodynamic starting time of the
injection was extrapolated from the average jet penetration length,
determined from sets of 200 images. The shadow images formed by source light
interaction with the spray were subsequently imaged to an interframe transfer
CCD array (PCO integrated by LaVision), at approximately 7:1 magnification.
Each frame was exposed to light from a single source laser pulse such that the
100~fs duration of each pulse ensured the motion of the jet and laboratory
seeing conditions were effectively frozen in each image. The minimum time
interval between the first and second frame in the image pair was limited to
$\sim$150~ns by the frame transfer time of the CCD. The resolution of the
complete imaging system was on the order of 7~$\mu$m.

\begin{figure}[h]
  \includegraphics[scale=0.5]{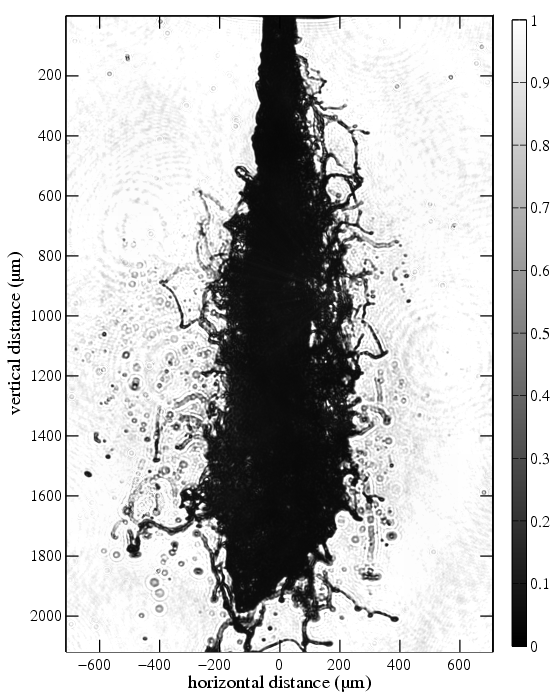}
  \caption{Shadow image of a diesel jet issuing from a 113~$\mu$m plain orifice
test injector into ambient air, 18~$\mu$s after SOI; $P_{inj}=60$~MPa.}
\label{fig:shadowimage}
\end{figure}

Using this experimental setup, pairs of images with precisely controlled time
separations were recorded. Figure~\ref{fig:shadowimage} shows an example image
for an injection pressure of 60~MPa. Here, the fuel jet can be seen issuing from
the tip of the injector located at the top of the frame and the diameter of the
injector orifice (113~$\mu$m) indicates the spatial scale.
Distinct liquid fragments and the edges of the jet which coincide with the
object plane appear sharply focused in the image. Structure and droplets near
the depth-of-field limits are also apparent but exhibit lower contrast and small
amounts of distortion consistent with defocusing. The signature of this
distortion allows out-of-focus regions to be excluded from the velocity analysis
\citep{Sedarsky2012_a}. Likewise, the large diffraction rings faintly visible in
the background of Fig.~\ref{fig:shadowimage} have negligible influence on
velocity results, since the focused features present higher signal levels
which dominate the correlation response.

\section{Velocity computation by image processing}
\label{analysis}
The calculations and data analysis discussed in the following section were
implemented in an in-house image analysis code based on OpenCV
\citep{Bradski2008} and developed for detailed examination of spray kinematics
and fluid motion.
By identifying persistent spatial intensity variations which are
present in consecutive images, the velocity of time-resolved liquid structures
in spray images can be estimated.
% The motion of time-resolved liquid structures in spray images can be
% estimated by matching persistent spatial intensity variations which are present in
% consecutive images.
%
This approach allows distinctive features in one image to be exploited by
correlation methods to track changes in one or more subsequent images which are
directly related to structure motion and the kinematics of the spray.

\subsection{Correlation approach}
\label{corrMethod}
\begin{figure}[h]\centering
  \includegraphics[scale=1]{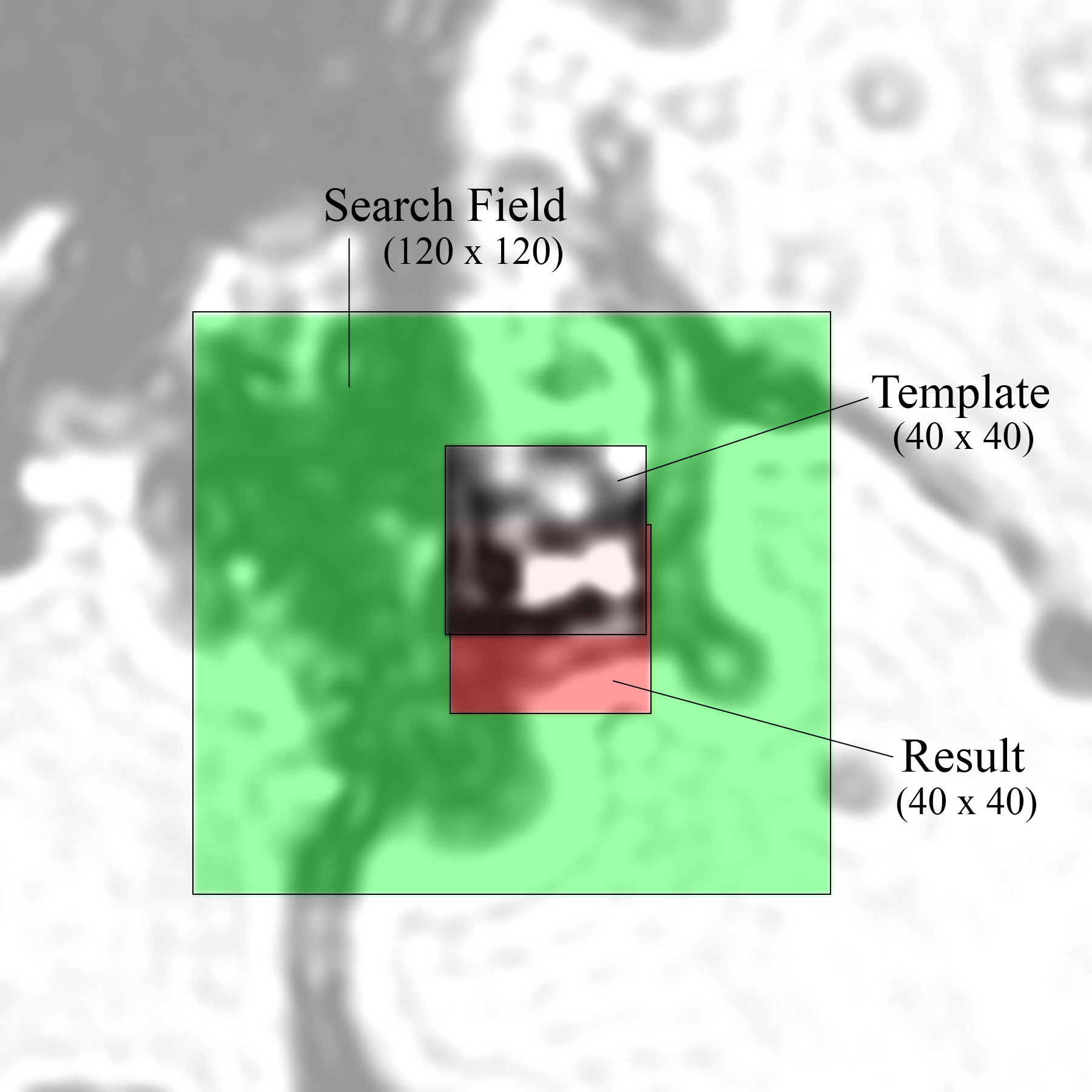}
  \caption{Image subregions for correlation matching analysis. Each vector, as
  shown in Fig.~\ref{fig:vectors}, represents the displacement of a sample
  region (Template) to it's best equivalent match (Result) within the Search
  Field region.}
  \label{fig:windows}
\end{figure}

On a basic level, the image data present a discretely sampled spatial intensity
pattern which indicates the underlying fluid structure. By considering the
amplitude and texture of the pattern within separately chosen spatial extents
(templates), localized feature sets can be formed from intensity data taken at
time, $t_1$. These templates are matched to the patterns within subregions
(search fields) chosen from subsequent intensity data taken at time, $t_2 = t_1
+ \Delta t$, yielding new spatial coordinates.
Assuming constant local image intensity, the displacement given by the
coordinate shift for each template/search pair indicates fluid motion over the
time interval, $\Delta t$.
Figure \ref{fig:windows} shows an example of one such set of image subregions,
together with the indicated match result.

Each template region is matched within its corresponding search field by
identifying the peak of the normalized cross correlation, given by:

\begin{equation}
\label{eqn:ncc}
\frac{1}{n-1}\sum_{x,y}\frac{1}{\sigma_S \sigma_T}(S(x,y)-\bar{S})(T(x,y)-\bar{T})
\end{equation}
where x and y represent horizontal and vertical image coordinates. $T(x,y)$ is
the template region, $S(x,y)$ is the search field, $n$ is the sum of the samples
(pixels) from the search and template regions, $\sigma_S$ and $\sigma_T$
represent the standard deviation of $S(x,y)$ and $T(x,y)$, and $\bar{T}$ and
$\bar{S}$ represent the respective average intensities of the template and search regions.

Thus, the general procedure for obtaining accurate velocity vectors from the raw
image data consists in: preparing image data, selecting template locations,
choosing search and template region sizes, and finally matching and validating
correlation results.

\subsection{Data preparation and sampling}
To minimize the relative error in the structure motion calculation, the
inter-frame time separation, $\Delta t$, was set to allow substantial
displacement of tracking features with negligible distortion in the intensity structure.
A time delay on the order of 260~ns was found to work well for following spray
motion generated by the current system (see Table~\ref{table:injectionparams}).

The correlation method given by Eqn.~\ref{eqn:ncc} implicitly requires
that the tracked features are minimally distorted between measurements, and
undergo translation, with negligible rotation, expansion, contraction, and
occlusion. This sets an upper limit on the acceptable time-delay for
cross-correlation of independent spatial intensity measurements.
A lower limit of the acceptable time-delay is imposed by the spatial resolution
and sensitivity of the imaging system, as smaller motions increase detection
requirements along with relative measurement uncertainty. In the present case,
however,
% In practice, however, 
% considerations such as electronic timing jitter or detector preparation
% capabilities generally limit the minimum achievable delay time. The practical
the minimum time-delay for the system is limited well above this level by the
frame-transfer timing of the detector.

Raw image data is adjusted prior to correlation analysis to ensure that the
data is consistent with the assumptions necessary to interpret pattern movement
as displacement of the imaged structure. Specifically, we require the
image-pairs to be registered to the same coordinate space, have a constant local
intensity (flat field), and minimal distortion ($\Delta t$ is small relative
to the detected velocity).

Since the matching approach discussed here is sensitive to intensity
fluctuations over the image,
% in addition to the form of the matched features,
variations which are unrelated to the measured structure should be minimized to
avoid influencing the matching process. In order to account for non-uniform
illumination and sensor effects, the data collected for each image includes dark
and `no signal' frames generated under the same conditions. This extra data is
used to apply a flat-field intensity correction to each image
\citep{Newberry1991}. All images are generated using the same source and
detector, so image-pairs are spatially synchronized without the need for image
registration or adjustments.

The cross-correlation of patterns formed by distributed structures with limited
variability can yield correlation coefficient topology with no clear maxima,
leading to displacement results which do not reflect the motion of the
underlying structure. This problem was addressed by identifying image regions
with significant variation and preferentially selecting these regions to form
the templates for correlation analysis. Target pixels used to locate the centers
of template regions were selected by thresholding the normalized image data and
applying Canny edge detection \citep{Lou2008_a} to identify image coordinates
near strong intensity gradients.

\subsection{Window sizing}
Accurate matching results depend on the identification of trackable image
regions and the selection of correlation window sizes which are suited to both
the spatial scale of the matched features and the time-separation of
the image data to be correlated.
In most cases, effective template regions should be large enough to loosely
frame smaller structures of interest, with corresponding search field regions 2
to 3 times the template dimensions.
%for the spray data presented here, square regions with 40 to 60 pixel sides
It is worthwhile to note that normalized cross-correlation is computationally
expensive (scaling as $O(n*\log n)$) and smaller functional window sizes are
preferable. In addition, the combination of template and search field sizes
naturally limit the range of detectable motion for the cross-correlation application. In the
limit of small templates, matching entropy is low, possibly resulting in errors
from ambiguous feature matching. In the limit of small search field size,
maximum displacement is severely constrained. For large templates, displacement
is likewise heavily constrained. Large template regions can also limit the
spatial resolution of the estimated velocity.

% 
% *The process can be expensive, the 2D cross correlation function
% needs to be computed for every point in the image. Calculation of the cross
% correlation function is itself a N2 operation. Ideally the mask should be
% chosen as small as practicable.
% 
%  *In many image identification processes the mask may need to be rotated and/or
%  scaled at each position.
% 
%  *This process is very similar to 2D filtering except in that case the image is
%  replaced by an appropriately scaled version of the correlation surface.

The optimal template and search field sizes depend on the size, form,
and intensity signature of the features of interest as well as the motion of the
underlying structure and time-separation of the image data. Window sizes which
are well-suited to the underlying data increase the efficiency of feature
identification and reduce the likelihood of erroneous pattern matching.

\subsection{Vector validation procedures}
% Relative criteria:
% *Correlation strength threshold
% *texture energy threshold
% *imposed physical likelihood weighting
% *imposed correlation structure weighting
% 
% Singular state (on/off):
% *sampling boundary constraints
% *imposed physical boundary constraints
% *stationary autocorrelation
% *reverse cross-correlation
% *outlier/statistical consistency
The cross-correlation operation for each template yields a match
result and displaced spatial coordinates which, together with the initial
template coordinates, represent a velocity for the identified feature set. With
properly sized correlation regions and significant variation in the template
pattern, this match result should indicate the motion of the underlying fluid
structure. However, a number of circumstances can result in matches which are
known to be, or likely to be false, such as edge cases, poorly structured
templates, or low coefficient matches.

\begin{figure}[h]\centering
  \includegraphics[scale=2.1]{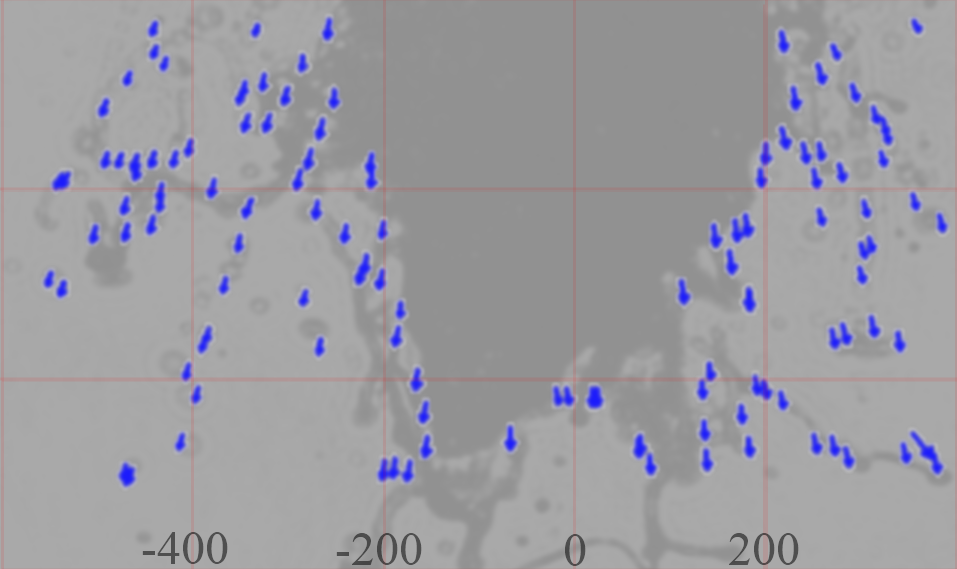}
  \caption{Velocity vectors of the leading edge of the spray shown in
  Fig.~\ref{fig:shadowimage}, with corresponding $\mu$m-scale spatial grid.}
  \label{fig:vectors}
\end{figure}

To identify and exclude these vectors, the velocity data and their associated
image sub-regions are validated against a list of criteria which examine the
spatial variance and texture energy of the image regions, as well as the
correlation strength and boundary constraints of the matching results.
The validation procedures fall roughly into two categories: threshold, or
`relative constraint' criteria, and absolute criteria. The threshold criteria
are user-selectable limits which can be set to eliminate weakly matching
patterns which are likely to yield erroneous vectors. Absolute criteria are
pass/fail tests, such as boundary contraints which eliminate vectors
indiscriminately.
The settings for effective validation do not vary widely, but as with
correlation window sizing, the optimum settings depend on the feature sets,
interframe delay, and underlying motion in the images.
The combination of the targeted selection of template regions and the threshold
validation criteria effectively limit the velocity results to in-focus edge
features of the spray \citep{Sedarsky2012_a}. Figure \ref{fig:vectors} shows
displacement vectors calculated from one image-pair near the leading edge
of the spray shown in Fig.~\ref{fig:shadowimage}.

It is important to point out that the results discussed here represent (2D)
planar motion of observed liquid structures---the projection of the real
structure (3D) velocity in the object plane of the imaging system.
Given the geometry of the plain-orifice jet, one can postulate that the third
(unknown) component of the velocity is on the order of the measured horizontal
velocity component.
In addition, it may be possible for the velocity of the observed structures
to differ from the velocity of the liquid itself, in which case the
quantitative application of these measurements would be limited.
Nevertheless, the kinematics presented here can be readily applied for
comparitive evaluation of high pressure injection conditions, and accurate
relative statistics on velocity can still be useful, especially for the
validation of numerical simulations.
\section{Statistical description of the injector}
\label{injector}
A statistical evaluation of the diesel test injector was conducted by compiling
the instantaneous velocities calculated from pairs of time-resolved images to
form data sets, fixed at different times relative to SOI.
Sets composed of 200 image-pairs were recorded and processed to obtain
instantaneous velocity fields. The vectors corresponding to these
image results (\x{2048}{2048}) were subsequently partitioned into
\x{20}{20} pixel cells.
In each of these square bins, the magnitudes of the computed vectors originating
from the region were averaged to yield a mean velocity magnitude assigned to
the cell.
The grid size was chosen to provide a reasonable compromise between sample
statistics and coarseness of the sample groups. In addition, the \x{20}{20} cell
size was small enough to permit permutation and averaging of the grid placement
allowing sub-grid resolution profiles, further increasing the accuracy of the
results.

The statistical relevance of the mean velocity for each region is directly
related to the number of vectors contributing to the block. On this basis, cells
containing less than 20 vectors were excluded from the analysis to improve the
significance of the computed results.
The velocity profiles derived from the data sets were arranged to form a
time-ordered series which was then used to chart spray behavior over the course
of the fuel injection event.

The targeting and validation procedures mentioned in section
\ref{analysis} ensure that out-of-focus elements of the image-pairs do not
contribute significantly to the velocity calculations. As a consequence, the
statistical velocity profiles exhibit a central region devoid of vectors which
corresponds roughly to the average jet position for each time-delay.

\begin{figure*}[htb]
\hspace{0.5cm}
  \begin{minipage}[b]{1\linewidth}
    \includegraphics[scale=0.30]{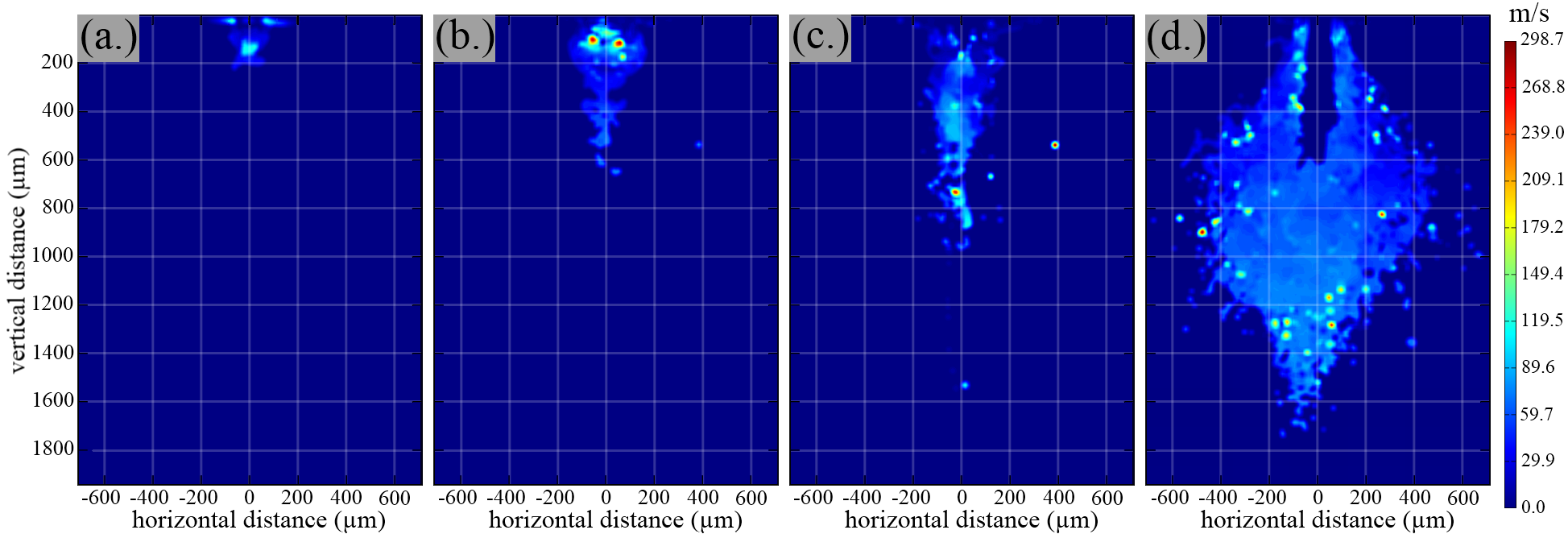}\vfill
  \end{minipage}
  \caption{Velocity profiles of the diesel test injector (see
table~\ref{table:injectionparams}), shown for a series of time-delays after SOI:
(a)~2~$\mu$s. (b)~5~$\mu$s. (c)~10~$\mu$s. (d)~15~$\mu$s.
  P$_{inj}$~=~60~MPa; Re~=~11~k; We~=~1400.}
  \label{fig:speed1}
\end{figure*}

\begin{figure*}[htb]
\hspace{0.5cm}
  \begin{minipage}[b]{1\linewidth}
    \includegraphics[scale=0.30]{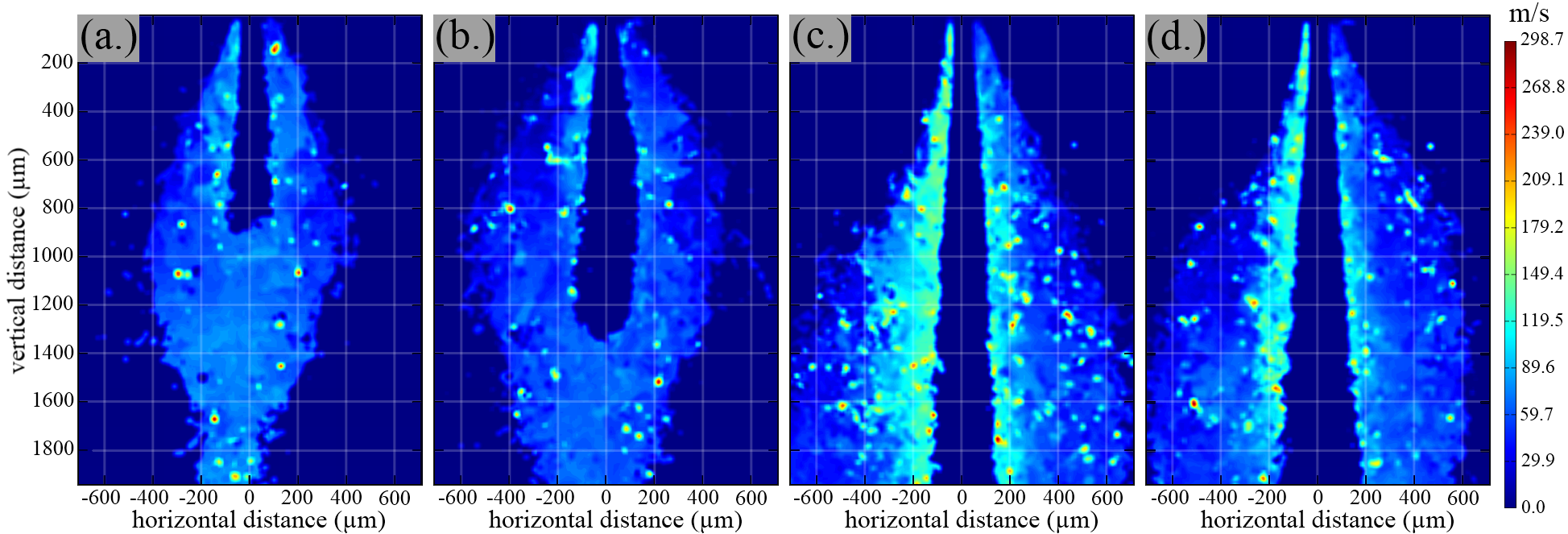}\vfill
  \end{minipage}
  \caption{Velocity profiles of the diesel test injector (see
table~\ref{table:injectionparams}), shown for a series of time-delays after SOI:
(a)~20~$\mu$s. (b)~25~$\mu$s. (c)~30~$\mu$s. (d)~35~$\mu$s.
  P$_{inj}$~=~60~MPa; Re~=~11~k; We~=~1400.}
  \label{fig:speed2}
\end{figure*}

\begin{figure*}[htb]
\hspace{0.5cm}
  \begin{minipage}[b]{1\linewidth}
    \includegraphics[scale=0.30]{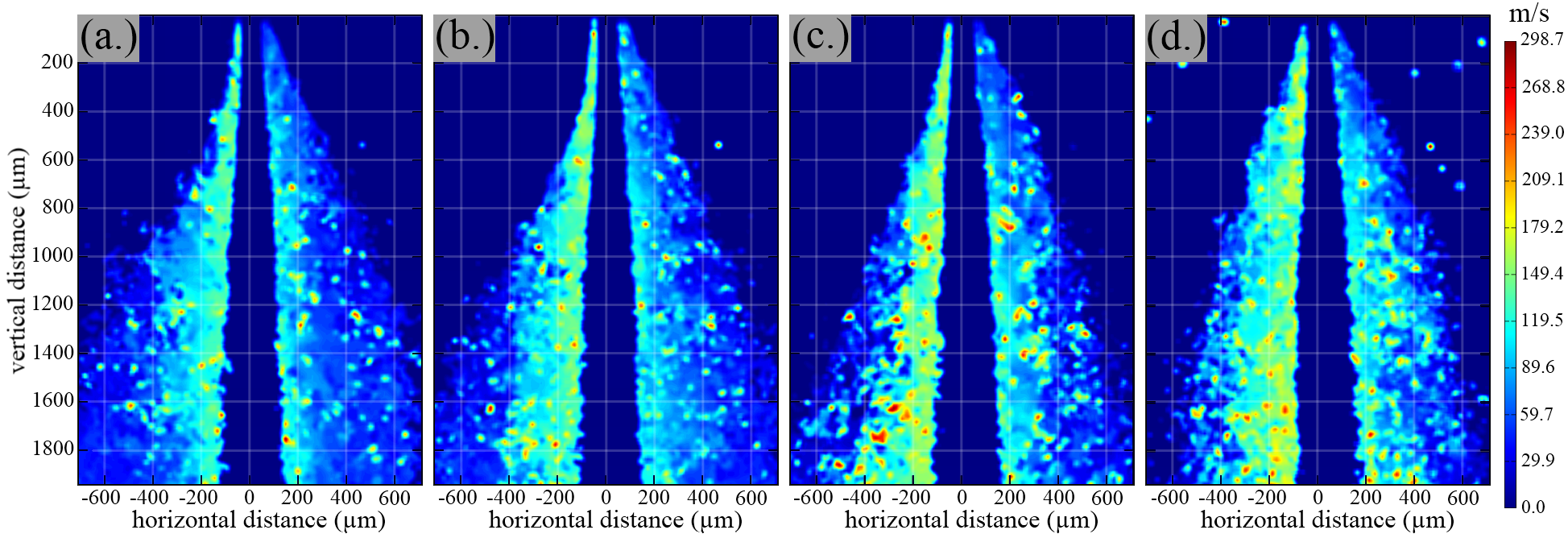}\vfill
  \end{minipage}
  \caption{Velocity profiles of the diesel test injector (see
table~\ref{table:injectionparams}), shown for a series of time-delays after
  SOI: (a)~40~$\mu$s. (b)~45~$\mu$s. (c)~50~$\mu$s. (d)~55~$\mu$s.
  P$_{inj}$~=~60~MPa; Re~=~11~k; We~=~1400.}
  \label{fig:speed3}
\end{figure*}

Figures~\ref{fig:speed1}, \ref{fig:speed2}, and \ref{fig:speed3} show the mean
velocity as a function of the location for different delays from the start of injection.
Examining figure~\ref{fig:speed1}, at very early time-delays we observe the
expected velocity profile, with the early jet extending centrally and slightly
slower velocities distributed on the portions of the jet with more lateral
motion.  For the first 10~$\mu$s after SOI, the flow is reasonably symmetric,
and the velocity profile initially appears uniformly distributed about the nozzle
orifice. Figure~\ref{fig:speed1} shows this uniform behavior with only small
speed differences visible between the left (fast) and right (slow) sides of the
spray at 15~$\mu$s after SOI.

As the spray becomes more established, however, a disparity in the velocity
profile becomes apparent, as evidenced by the progression from 20 to 35~$\mu$s
shown in Fig.~\ref{fig:speed2}.
The evolution of the spray continues with the fast and slow sides of the
jet becoming more pronounced, reaching differentials of 90 m/s by 40~$\mu$s.
This asymmetric profile continues to intensify as the spray develops.
Increasing speeds can be observed on the slow side of the jet profile as well,
as the jet expands downstream. Nevertheless, strong differences in velocity across the jet
profile are apparent through the entire range as the flow evolves, before
reaching a steady flow condition at around 55~$\mu$s after SOI.

\begin{figure}[h]\centering
  \includegraphics[scale=1.1]{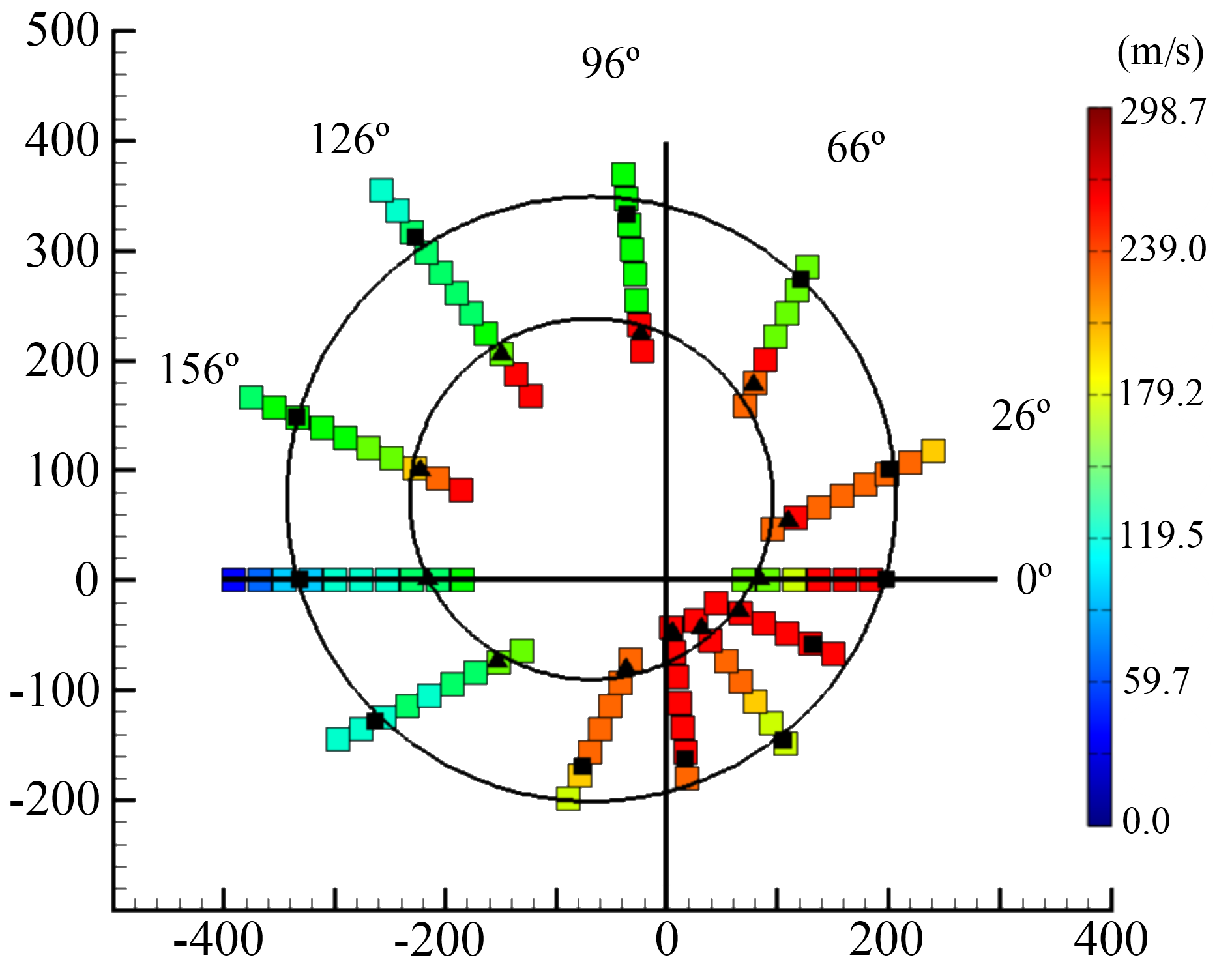}
  \caption{Top-down view of velocity profiles the for single-hole diesel test
  nozzle, showing results compiled from image data and velocity analysis of the
  spray at a range of angles. The spatial scale is given in $\mu$m, and the
  colorbar represents velocity magnitude in units of m/s. The inner and outer
  circles are fit to the edges of the jet 1~mm downstream from the
  nozzle orifice. The edges are calculated from spray images averaged over 200
  shots and detected at the 20\% and 80\% threshold levels to show the
  variability and extent of the spray.
  }
\label{fig:angles}
\end{figure}

In order to understand the shape of the liquid jet, the injector was rotated
around the vertical axis and sets of 200 image-pairs were acquired at 6
different angles for additional velocity analysis.
Figure~\ref{fig:angles} shows a map of the velocity field from liquid structure
motion for angles of 0$^{\circ}$, 26$^{\circ}$, 66$^{\circ}$,
96$^{\circ}$, 126$^{\circ}$ and 156$^{\circ}$ chosen from an arbitrary reference. These data were calculated
from the fully developed spray, for a time-delay $\sim$55~$\mu$s after SOI.
These measurements confirm that the present injector
produces an asymmetric velocity profile, which may be a consequence of
cavitation and irregular flow inside the injector.
Note that this behavior is persistent and reproducible;
the distinct fast and slow sides of the injector remain the same for every
injection event.

In order to reconstruct the shape of a section of the spray, the diameter of the
jet at a position 1~mm downstream from the nozzle orifice is plotted for
different nozzle viewing angles in Fig.~\ref{fig:angles}. The inner and outer
circles shown in Fig.~\ref{fig:angles} delineate the average periphery of the spray
derived from 200 images for each angle.

Given the shot-to-shot variability in the form and motion of the jet, the
calculated location of the outer edge of the spray is sensitive to the choice
of gray-level threshold used in the edge-finding algorithm. To show the
range of this variability, two levels are shown in the figure, where the inner
and outer lines are fit to data calculated with 20\% and the 80\% threshold
levels, respectively. These lines show approximately circular cross-sections,
with centers which are shifted from the center of the nozzle orifice in the
direction of the low-velocity side of the spray.
% This result is consistent with the spray images which show a deflection of the
% liquid jet in the same direction.

The colored lines radiating from the orifice position and indicated by the
angle labels in Fig.~\ref{fig:angles} show the velocity magnitude profile for
each rotation angle. The distribution of velocity in the profiles clearly
indicates distinct fast and slow sides of the spray, approximately centered at
126$^\circ$. The velocity differences apparent in this angular view are
consistent with the deflection of the liquid jet which is visible in the spray
images. This deflection angle was estimated from the spray periphery data
%shown in Fig.~\ref{fig:angles} and determined
to be $\sim$5.5$^\circ$. 

The lack of symmetry between the velocity profiles for each angle
implies that the out-of-plane component may be significant for
angles which are not aligned to the side of the jet, i.e. direct views of the
fast or slow sides of the jet, such as the profile for 26$^\circ$ shown in
Fig.~\ref{fig:angles}.
This is important for effective use of the velocimetry approach applied in this
work, which measures velocity confined to the object plane of the imaging setup
and is unable to determine the out-of-plane component of the structure motion.

\section{Atomization behavior}
\label{atomization}
\begin{figure}[h]\hspace{-1.2cm}
  \includegraphics[scale=0.38]{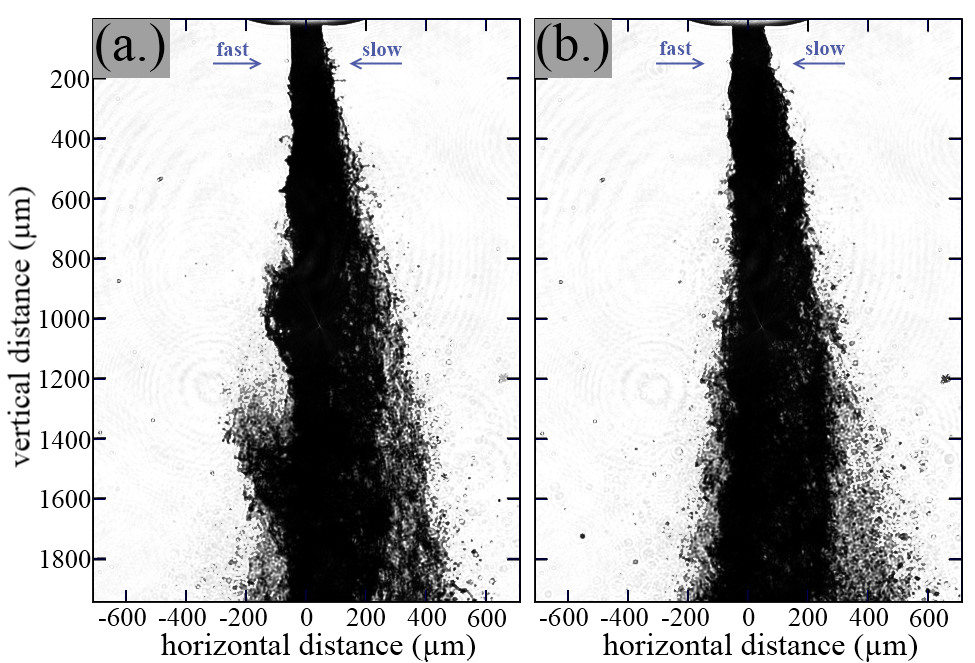}
  \caption{Example images showing the 126$^\circ$ view (see
  Fig.\ref{fig:angles}) of the test injector spray. Two
  different atomization processes are apparent on the left side of the liquid column,
  which corresponds to the highest liquid velocity. $55 \mu$s after
  SOI; $P_{inj}=60$~MPa.}
\label{fig:breakup}
\end{figure}

The large velocity differences observed in the liquid structures on opposing
sides of the jet accompany changes in the prevailing atomization conditions
across the spray.
Figure \ref{fig:breakup} shows example images of the fully developed
spray viewed from the 126$^{\circ}$ position, as shown in
Fig.~\ref{fig:angles}. Here, the left-hand edge (fast side) of the jet exhibits
the highest average liquid structure velocities while the right-hand edge
(slow side) directly opposite exhibits the lowest average velocities.

As noted above, the trajectory of the liquid column exhibits
an angular deflection in the direction of the slow side of the nozzel outlet.
This deflection extends the slow edge laterally as the spray is
established, increasing the exposed shear surface and perturbing boundary layers
forming at the spray edge. These are small effects under the current
conditions, but they contribute to the asymmetric form of the flow and the
spread of the droplets distributed on the slow side of the jet, as seen on
the right side of both images shown Fig.~\ref{fig:breakup}.

The slow side of the jet begins to breakup  in the
vicinity of the nozzle within 1 or 2 nozzle diameters. The jet surface near
the orifice is rough as it exits, including small disturbances which contribute
to momentum exchange, entrainment with the sourrounding air, and shearing of
liquid from the jet surface.

The character of the atomization on the fast side of the jet differs
significantly from the slow side kinematics. At first, the trajectory of the
liquid along the edge curves inward and on average, droplets are dispersed at a
greater distance from the nozzle. The jet surface near the orifice appears
undisturbed and a smooth liquid column extends, in most cases, for 2 to 5 nozzle
diameters before air entrainment and shear effects begin to disturb the jet.

Distinct macrostructure appears on the fast side of the spray under the
asymmetric velocity conditions. In about half the cases, the breakup on the fast
side of the spray resembles the slow edge, with shear forces driving
Kelvin--Helmholtz instabilities which grow to form ligaments and small droplets.
However, breakup on the slow edge happens more rapidly, with droplets and
structure appearing earlier and with more intricate interaction.
In other cases, small instabilities become apparent on the fast side within the
first few nozzle diameters and quickly grow to produce large waves which
periodically shed larger droplets and ligaments.
This behavior is intermittent and present throughout the range of injection
pressures available with the current system. Figure~\ref{fig:breakup} includes
two example images from the same data set which show both of these breakup modes
for the same view angle and identical conditions.

Reproducible asymmetric needle motion has been observed in diesel fuel injectors
and shown to contribute to flow irregularities \citep{Powell2011}.
However, the flow velocity differences and transient atomization modes with periodic
structure observed here resemble flow structures which are often seen in the
presence of cavitation produced under controlled conditions in scaled,
transparent test nozzles. This likeness suggests that the observed flow
conditions are the result of cavitation and fluctuating pressure conditions in
the fluid upstream of the nozzle orifice.

\begin{figure}[h]
  \includegraphics[scale=2]{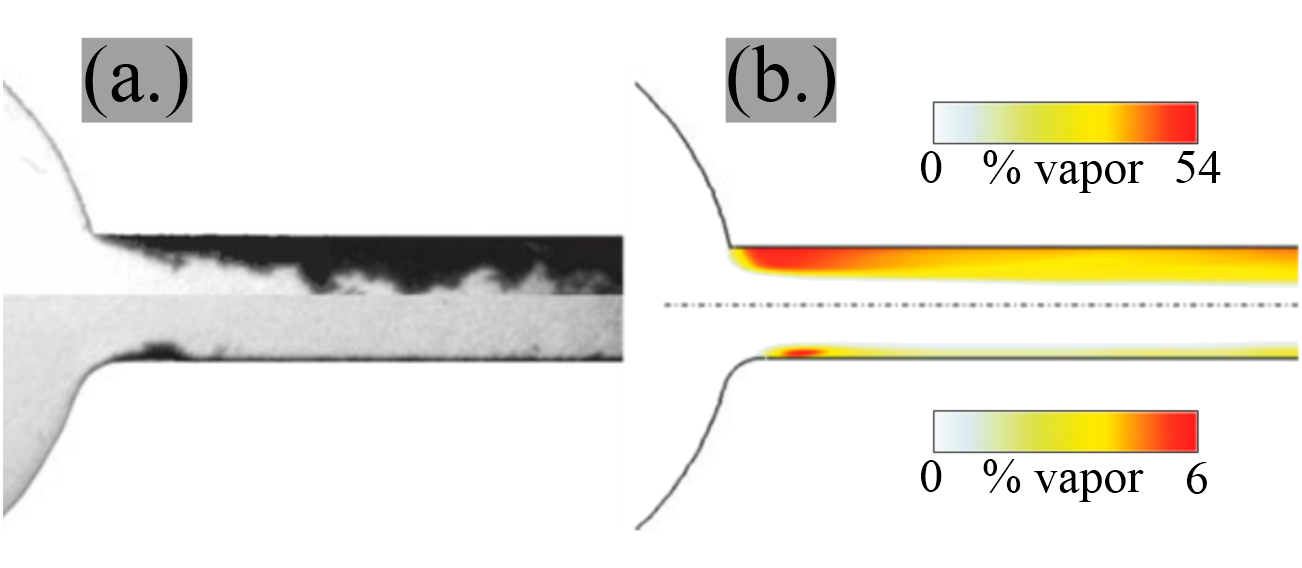}
  \caption{Half-nozzle diesel injection (P$_{inj}=50$~MPa, Re$\approx
  19$k) images and simulations for sharply
  edged (top) and hydroground nozzles (bottom). Part (a) shows
  time-resolved shadowgrams of cavitation vapor from
  K\"onig \& Blessing (2000). Part (b) shows simulation results of mean
  cavitation vapor distribution (volume percentage) from Giannadakis, et al. (2008).}
\label{fig:Giannadakis}
\end{figure}

Studies of cavitating flows show that viscous stress and pressure effects
largely determine the inception of cavitation within the channel upstream of the
orifice \citep{Dabiri2007,Dabiri2010}. Here, vortices formed near the inlet can
lead to local pressure and temperature conditions which induce the production of
vapor from the bulk liquid. The formation and subsequent collapse of bubbles in
the fluid give rise to pressure waves which can interact to form oscillatory
behavior and instabilities which are apparent in the emerging spray
\citep{Mauger2012thesis}.
Figure~\ref{fig:Giannadakis} shows measurements and simulation results
from previous work \citep{Giannadakis2008} comparing a single-hole
injector flow with sharply edged (top) and hydroground nozzles (bottom) for
flow conditions comparable to test injector presented in this work.

The shadowgrams of the nozzle interior shown in figure~\ref{fig:Giannadakis}(a)
give a qualitative view of the formation and propagation of cavitation in the
channel.
Figure~\ref{fig:Giannadakis}(b) shows the amount and spatial extent of cavitation vapor predicted
by the simulations of Giannadakis for these flow conditions.

Note that the nozzle used in the present work is operated at a lower Reynolds
number than the flows presented in Fig.~\ref{fig:Giannadakis}, and the nozzle
construction corresponds to the cases shown in the bottom half of
figs.~\ref{fig:Giannadakis}(a) and \ref{fig:Giannadakis}(b).
Consequently, modest amounts of cavitation and a symmetric general profile are
expected in the flow according to the test injector design. Nevertheless, strong
differences are apparent between fast side and slow sides of the spray, and
intermittent changes in the morphology of the spray under identical conditions
indicate that transient phenomena are active in the formation of the spray. 

X-ray microscopy images of the injector with $\sim$20~$\mu$m resolution (see
Fig.~\ref{fig:injector}) were taken for a range of angles about the central
axis. These images indicate a symmetric nozzle construction with no visible
irregularities. However, it is possible that a material defect or error in the
machining process has left an uneven region at the channel inlet which was not
covered by the microscopic inspection.
Gross differences in velocity and deflection have been observed in single-hole
sprays in some conditions when a portion of the inlet lip is lower (exhibiting a
different curvature) than the surrounding edge. This can lead to vortex
interactions which encourage midstream or string cavitation along one side of
the channel, reducing the effective area of the orifice and creating asymmetric
flow conditions \citep{Danckert2001}.

The onset of the velocity disparity across the injector is consistent with a
cavitation or dynamic pressure condition, where small pockets of vapor form as
a critical velocity is reached some time $\sim$10~$\mu$s after SOI. Pockets of
gas forming in the flow could then serve to isolate the liquid from the wall of
the channel, disturbing the boundary layer and giving rise to turbulence in the
internal flow. This in turn would promote small scale instabilities which
contribute to the prevalence of different atomization conditions on each side
of the spray.

Even when vortex effects do not produce cavitation directly, pressure
fluctuations set up in the channel as a result of the interaction can contribute
to spray instabilities and periodic behavior appearing in the downstream regions
of the flow.
It is likely that the large oscillatory features appearing on the fast edge of
the diesel spray are the result of pressure fluctuation dynamics related to
cavitation which grow to destabilize the liquid column as they propagate
downstream.

%combination of pressure and kinetic energy

\section{Conclusions}
A statistical description of a single-hole diesel fuel injection spray,
including high-resolution velocity information, was compiled from ultrafast
imaging measurements.

The time-resolved spray measurements were made with an imaging system
synchronized to a dual-pulse femtosecond laser source, allowing the
acquisition of time-correlated image-pairs which are spatially resolved by the
optics. Correlation analysis was successfully applied to these image results to
calculate the instantaneous spray velocity related to the time-step for each image-pair.

This approach is able to effectively map the early time-evolution of the spray.
In the case of the current injector, it reveals large disparities in the
liquid structure velocity in different regions across the spray.

Given the flow conditions and the construction of the nozzle studied here, it is
likely that the transient atomization behavior and asymmetric flow conditions
exhibited by the test injector are the result of an irregularity in the nozzle
inlet, causing cavitation and asymmetric pressure fluctuations within the nozzle
itself. This conclusion is consistent with the form of the observed
breakup modes, the deflection of the fuel jet, and would explain the asymmetric
velocity behavior measured by the ultrafast ICV statistics.

Statistically significant collections of spray data compiled from these spatial- and
temporally resolved single-shot measurements provide an effective and
relatively straightforward view of spray structure velocities which are relevent
to primary breakup and atomization in multiphase flows.

% Sets of time-resolved image data were acquired at a series of injection
% pressures and times following the start of injection, and analyzed
% to obtain instantaneous velocities of spray structures.
% 
% A statistical description of the diesel spray atomization was compiled from
% these spatial- and temporally resolved single-shot measurements, providing an
% effective and relatively straightforward view of spray structure velocities
% relevent to primary breakup and flow conditions in the spray.

% \begin{figure*}[bht]
% \centering
%   \begin{minipage}[b]{1\linewidth}
%     \includegraphics[scale=0.45]{Speed4.png}\vfill
%   \end{minipage}
%   \caption{Empty caption, but the paper is still really cool.}
%   \label{fig:speed4}
% \end{figure*}

% \begin{figure*}[th]
% \centering
%   \begin{minipage}[b]{1\linewidth}
%     \includegraphics[scale=0.32]{Speed4-1.png}\\
%     \includegraphics[scale=0.32]{Speed4-2.png}\\
%     \includegraphics[scale=0.32]{Speed4-3.png}
%   \end{minipage}
%   \caption{Spray velocity profiles for single-hole diesel test nozzle
%   for injection pressure, $P_{inj}=40$~MPa. The speed (averaged over 200 shots)
%   of time-resolved liquid structure is shown versus delay from the start of
%   injection (SOI). The spatial scale is given in $\mu$m, and the colorbar
%   represents velocity magnitude in the object plane in units of m/s.}
%   \label{fig:results}
% \end{figure*}

\begin{acknowledgements}
This work was supported by NADIA-Bio program, with funding from the French
Government and the Haute-Normandie region, in the framework of the Moveo cluster
(``private cars and public transport for man and his environment").
\end{acknowledgements}

\bibliographystyle{abbrvnat}
\bibliography{sg-references}      % name your BibTeX data base

\begin{thebibliography}{27}
\providecommand{\natexlab}[1]{#1}
\providecommand{\url}[1]{\texttt{#1}}
\expandafter\ifx\csname urlstyle\endcsname\relax
  \providecommand{\doi}[1]{doi: #1}\else
  \providecommand{\doi}{doi: \begingroup \urlstyle{rm}\Url}\fi

\bibitem[Bachalo(1994)]{Bachalo1994}
W.~D. Bachalo.
\newblock Experimental methods in multiphase flows.
\newblock \emph{International Journal of Multiphase Flow}, 20\penalty0
  (0):\penalty0 261 -- 295, 1994.

\bibitem[Bradski and Kaehler(2008)]{Bradski2008}
G.~Bradski and A.~Kaehler.
\newblock \emph{Learning OpenCV: computer vision with the OpenCV library}.
\newblock O'Reilly Media, 1 edition, Sept. 2008.
\newblock ISBN 978-0-596-51613-0.

\bibitem[Chaves et~al.(2004)Chaves, Kirmse, and Obermeier]{Chaves2004}
H.~Chaves, C.~Kirmse, and F.~Obermeier.
\newblock Velocity measurements of dense diesel fuel sprays in dense air.
\newblock \emph{Atomization and Sprays}, 14\penalty0 (6), 2004.

\bibitem[Dabiri et~al.(2007)Dabiri, Sirignano, and Joseph]{Dabiri2007}
S.~Dabiri, W.~A. Sirignano, and D.~D. Joseph.
\newblock Cavitation in an orifice flow.
\newblock \emph{Physics of Fluids}, 19\penalty0 (7):\penalty0
  072112.1--072112.9, 2007.

\bibitem[Dabiri et~al.(2010)Dabiri, Sirignano, and Joseph]{Dabiri2010}
S.~Dabiri, W.~A. Sirignano, and D.~D. Joseph.
\newblock A numerical study on the effects of cavitation on orifice flow.
\newblock \emph{Physics of Fluids}, 22\penalty0 (4):\penalty0
  042102--042102--13, 2010.

\bibitem[Danckert and Affolter(2001)]{Danckert2001}
B.~Danckert and P.~K. Affolter.
\newblock Ways of nozzle geometry optimisation for new injection systems,
  fulfilling future emission limits.
\newblock In \emph{Proc. of the 23rd World Congress on Combustion Engine
  Technology for Ship Propulsion, Power Generation, Rail Traction}, volume~2 of
  \emph{Proceedings of CIMAC}. CIMAC, May 2001.

\bibitem[Fielding et~al.(2001)Fielding, Long, Fielding, and
  Komiyama]{Fielding2001}
J.~Fielding, M.~B. Long, G.~Fielding, and M.~Komiyama.
\newblock Systematic errors in optical-flow velocimetry for turbulent flows and
  flames.
\newblock \emph{Applied Optics}, 40\penalty0 (6):\penalty0 757--764, Feb. 2001.

\bibitem[Giannadakis et~al.(2008)Giannadakis, Gavaises, and
  Arcoumanis]{Giannadakis2008}
E.~Giannadakis, M.~Gavaises, and C.~Arcoumanis.
\newblock Modelling of cavitation in diesel injector nozzles.
\newblock \emph{Journal of Fluid Mechanics}, 616:\penalty0 153--193, 2008.

\bibitem[Hespel et~al.(2012)Hespel, Blaisot, Gazon, and Godard]{Hespel2012}
C.~Hespel, J.-B. Blaisot, M.~Gazon, and G.~Godard.
\newblock Laser correlation velocimetry performance in diesel applications:
  spatial selectivity and velocity sensitivity.
\newblock \emph{Experiments in Fluids}, 53\penalty0 (1):\penalty0 245--264,
  July 2012.

\bibitem[Idlahcen et~al.(2012)Idlahcen, Roz{\'{e}}, Girasole, and
  Blaisot]{Idlahcen2012}
S.~Idlahcen, C.~Roz{\'{e}}, T.~Girasole, and J.-B. Blaisot.
\newblock Sub-picosecond ballistic imaging of a liquid jet.
\newblock \emph{Experiments in Fluids}, 52\penalty0 (2):\penalty0 289--298,
  2012.

\bibitem[Kristensson et~al.(2010)Kristensson, Richter, and
  Ald{\'{e}}n]{Kristensson2010}
E.~Kristensson, M.~Richter, and M.~Ald{\'{e}}n.
\newblock Nanosecond structured laser illumination planar imaging for
  single-shot imaging of dense sprays.
\newblock \emph{Atomization and Sprays}, 20\penalty0 (4):\penalty0 337--343,
  Sept. 2010.

\bibitem[Kr{\"{u}}ger and Gr{\"{u}}nefeld(1999)]{Kruger1999}
S.~Kr{\"{u}}ger and G.~Gr{\"{u}}nefeld.
\newblock Stereoscopic flow-tagging velocimetry.
\newblock \emph{Applied Physics B: Lasers and Optics}, 69\penalty0
  (5):\penalty0 509--512, 1999.

\bibitem[Lebas et~al.(2009)Lebas, M{\'{e}}nard, Beau, Berlemont, and
  Demoulin]{Lebas2009}
R.~Lebas, T.~M{\'{e}}nard, P.~A. Beau, A.~Berlemont, and F.~X. Demoulin.
\newblock Numerical simulation of primary break-up and atomization: Dns and
  modelling study.
\newblock \emph{International Journal of Multiphase Flow}, 35\penalty0
  (3):\penalty0 247 -- 260, 2009.

\bibitem[Linne et~al.(2009)Linne, Paciaroni, Berrocal, and Sedarsky]{Linne2009}
M.~A. Linne, M.~E. Paciaroni, E.~Berrocal, and D.~Sedarsky.
\newblock Ballistic imaging of liquid breakup processes in dense sprays.
\newblock \emph{Proc. of the Combustion Institute}, 32:\penalty0 2147--2161,
  2009.

\bibitem[Luo and Duraiswami(2008)]{Lou2008_a}
Y.~Luo and R.~Duraiswami.
\newblock Canny edge detection on nvidia cuda.
\newblock In \emph{Computer Society Conference on Computer Vision and Pattern
  Recognition Workshops}, Proceedings of CVPRW, pages 1--8. IEEE, june 2008.

\bibitem[Marks et~al.(2010)Marks, Hershey, and Movellan]{Marks2010}
T.~K. Marks, J.~R. Hershey, and J.~R. Movellan.
\newblock Tracking motion, deformation, and texture using conditionally
  gaussian processes.
\newblock \emph{Pattern Analysis and Machine Intelligence, IEEE Transactions
  on}, 32\penalty0 (2):\penalty0 348 --363, feb. 2010.

\bibitem[Mauger(2012)]{Mauger2012thesis}
C.~Mauger.
\newblock \emph{Cavitation dans un micro-canal mod{\`{e}}le d’injecteur
  diesel : m{\'{e}}thodes de visualisation et influence de l’{\'{e}}tat de
  surface}.
\newblock PhD thesis, University of Lyon, Apr. 2012.

\bibitem[M{\'{e}}nard et~al.(2007)M{\'{e}}nard, Tanguy, and
  Berlemont]{Menard2007}
T.~M{\'{e}}nard, S.~Tanguy, and A.~Berlemont.
\newblock Coupling level set/vof/ghost fluid methods: Validation and
  application to 3d simulation of the primary break-up of a liquid jet.
\newblock \emph{International Journal of Multiphase Flow}, 33\penalty0
  (5):\penalty0 510 -- 524, 2007.

\bibitem[Newberry(1991)]{Newberry1991}
M.~V. Newberry.
\newblock Signal-to-noise considerations for sky-subtracted ccd data.
\newblock \emph{Publications of the Astronomical Society of the Pacific},
  103\penalty0 (659):\penalty0 122--130, 1991.

\bibitem[Raffel et~al.(2007)Raffel, Willert, Wereley, and
  Kompenhans]{Raffel2007}
M.~Raffel, C.~E. Willert, S.~T. Wereley, and J.~Kompenhans.
\newblock \emph{Particle image velocimetry : a practical guide}.
\newblock Springer, New York, 2 edition, 2007.
\newblock ISBN 9783540723073.

\bibitem[Ram{\'{\i}}rez et~al.(2009)Ram{\'{\i}}rez, Som, Aggarwal, Kastengren,
  El-Hannouny, Longman, and Powell]{Ramirez2009}
A.~Ram{\'{\i}}rez, S.~Som, S.~Aggarwal, A.~Kastengren, E.~El-Hannouny,
  D.~Longman, and C.~Powell.
\newblock Quantitative x-ray measurements of high-pressure fuel sprays from a
  production heavy duty diesel injector.
\newblock \emph{Experiments in Fluids}, 47\penalty0 (1):\penalty0 119--134,
  2009.

\bibitem[Sedarsky et~al.(2006)Sedarsky, Paciaroni, Linne, Gord, and
  Meyer]{Sedarsky2006}
D.~Sedarsky, M.~E. Paciaroni, M.~A. Linne, J.~R. Gord, and T.~R. Meyer.
\newblock Velocity imaging for the liquid-gas interface in the near field of an
  atomizing spray: proof of concept.
\newblock \emph{Opt. Lett.}, 31\penalty0 (7):\penalty0 906--8, 2006.

\bibitem[Sedarsky et~al.(2009)Sedarsky, Gord, Carter, Meyer, and
  Linne]{Sedarsky2009}
D.~Sedarsky, J.~R. Gord, C.~Carter, T.~R. Meyer, and M.~A. Linne.
\newblock Fast-framing ballistic imaging of velocity in an aerated spray.
\newblock \emph{Opt. Lett.}, 34\penalty0 (18):\penalty0 2748--2750, 2009.

\bibitem[Sedarsky et~al.(2011)Sedarsky, Berrocal, and Linne]{Sedarsky2011}
D.~Sedarsky, E.~Berrocal, and M.~A. Linne.
\newblock Quantitative image contrast enhancement in time-gated
  transillumination of scattering media.
\newblock \emph{Opt. Express}, 19\penalty0 (3):\penalty0 1866--1883, Jan 2011.

\bibitem[Sedarsky et~al.(2012)Sedarsky, Idlahcen, Blaisot, and
  Roz{\'{e}}]{Sedarsky2012_a}
D.~Sedarsky, S.~Idlahcen, J.-B. Blaisot, and C.~Roz{\'{e}}.
\newblock Planar velocity analysis of diesel spray shadow images.
\newblock In \emph{Proc. of multiphase flow and transport phenomena}, MFTP,
  Apr. 2012.

\bibitem[Shavit and Chigier(1995)]{Shavit1995}
U.~Shavit and N.~Chigier.
\newblock Fractal dimensions of liquid jet interface under breakup.
\newblock \emph{Atomization and Sprays}, 5\penalty0 (6):\penalty0 525--543,
  1995.

\bibitem[Tokumaru and Dimotakis(1995)]{Tokumaru1995}
P.~T. Tokumaru and P.~E. Dimotakis.
\newblock Image correlation velocimetry.
\newblock \emph{Experiments in Fluids}, 19:\penalty0 1--15, 1995.

\end{thebibliography}

%\begin{thebibliography}{sg-references}
% %
% % and use \bibitem to create references. Consult the Instructions
% % for authors for reference list style.
% % Format for Journal Reference
% Author, Article title, Journal, Volume, page numbers (year)
% Author, Book title, page numbers. Publisher, place (year)
%\end{thebibliography}
%\bibliography{sg-references}   % name your BibTeX data base
\end{document}